\documentclass[10pt]{article}

\begin{document}
\input amssym.tex

\title{Accelerated frames in de Sitter spacetime}
\author{Ion I. Cot\u aescu \thanks{E-mail:~cota@physics.uvt.ro; icotaescu@yahoo.com}\\
{\it West University of Timi\c soara,}\\{\it V. Parvan Ave. 4,
RO-300223 Timi\c soara}}
\maketitle 
\begin{abstract}
We propose a definition of the accelerated frames in de Sitter spacetime whose metric recovers the Rindler approach in the flat limit.  Unfortunately, this metric has no a satisfactory limit for vanishing acceleration such that its physical meaning remains obscure.  

Pacs:{04.62.+v}
\end{abstract}

\vspace*{2cm} 

\newpage

The Rindler metric of the accelerated frames in Minkowski spacetime can be formally related to a chart of  the de Sitter manifold \cite{BD, SS} whose metric is a conformal Rindler one.  Could we interpret this chart as being the typical one of an accelerated observer in the de Sitter  background ? We show that by generalising the Rindler transformation  between a static and an accelerated frame in this geometry we may obtain a new metric of the accelerated frame.

The physical meaning of the coordinate transformations in general relativity \cite{BD} can be investigated focusing on the transformation rules of the conserved quantities. This can be done  in a classical manner, analysing the transformation of the classical conserved quantities,  or from the quantum mechanical point of view, looking for the properties of the conserved operators, i.e. the Killing vector fields generating the natural representation of the isometry group.  We show that this last method helps us to generalize the Rindler approach \cite{MB,R} to the de Sitter spacetime.  

In the $(1+1)$-dimensional Minkowski spacetime the Rindler transformation 
\begin{eqnarray}
T&=&x \sinh(a t)\,,\\
X&=&x \cosh(a t)\,,
\end{eqnarray} 
between the static frame $(T,X)$ and the accelerated one $(t,x)$ allows one to find the line element 
\begin{equation}\label{line}
ds^2=dT^2-dX^2=a^2 x^2 dt^2-dx^2\,,
\end{equation} 
and the identity
\begin{equation}
aK=i a (X\partial_{T}+T\partial_{X})=i\partial_t\,,
\end{equation}
which shows that the energy operator of the accelerated frame, $H_{acc}=i\partial_t$, coincides to the generator of the Lorentz boosts in the static  frame $(T,X)$ multiplied with the acceleration $a$. 

We ask if one could define accelerated frames in the de Sitter spacetime starting with an analogous transformation between the energy operator and the generator of the Lorentz boosts which are well-defined in the de Sitter case too \cite{C1}. 
   
We consider the expanding portion of the $(1+1)$-dimensional de Sitter spacetime equipped with the  chart $(T,X)$ of conformal time $T\in(-\infty,0]$ and the line element 
\begin{equation}\label{mconf}
ds^{2}=\frac{1}{\omega^2 {T}^2}\left({dT}^{2}-d{X}^2\right)\,,
\end{equation}
where $\omega$ denotes the Hubble parameter of the de Sitter hyperboloid. The scalar quantum modes in this chart are determined as common eigenfunctions of the Klein-Gordon operator and one more conserved operator which must be an isometry generator of the natural representation. The $SO(1,2)$ isometries give rise to three operators with physical meaning namely, the energy, $H$, momentum, $P$, and the generator of the Lorentz boosts,$K$, that read \cite{C2}
\begin{eqnarray}
H&=&-i\omega(T\partial_{T}+X\partial_{X})\,, \\
P\,&=&-i\partial_{X}\,,\\
K&=&-i\omega X T\partial_{T}+\frac{i}{2\omega}[1-\omega^2(T^2+X^2)]\partial_{X}\,, \label{K}
\end{eqnarray}
The Killing vector field of the energy operator $H_1$ is timelike inside the light cone where $T^2>X^2$ \cite{C1} while that of $K_1$ is timelike in four diagonal rectangular sheets given by the condition $[\omega^2(T+X)^2-1][\omega^2(T-X)^2-1]<0$. In these domains  the operator $K_1$ could play the role of an energy operator.

Now we assume that, as in the Rindler case,  the observer in the accelerated frame $(t,x)$ measures the eigenvalues of  the energy operator $H_{acc}=aK$  instead of $H$. The problem is to determine the coordinate transformation between these frames with the natural initial conditions
\begin{equation}
\left.T(t,x)\right |_{t=0}=-\frac{1}{\omega}\,, \quad \left. X(t,x)\right|_{t=0}=x\,,
\end{equation}
since $T$ is a conformal time \cite{BD}. 

This problem can be solved assuming that the coordinates $ t,\, x$ are of the de Sitter-Painlev\' e type \cite{G,P} so that $H_{acc}=i\partial_t$ \cite{Pascu}.  Then, after a few manipulations and by using suitable codes under Maple  we find that the unique solution is the following transformation:  
\begin{eqnarray}
T(t,x)&=&-\frac{2}{\omega \left[2+2\omega x\sinh(a t)+\omega^2 x^2(\cosh(a t)-1)\right]}\,,\\
X(t,x)&=&\frac{2x\cosh(a t)+\omega x^2 \sinh(a t)}{\omega \left[2+2\omega x\sinh(a t)+\omega^2 x^2(\cosh(a t)-1)\right]}\,.
\end{eqnarray}
Two properties are remarkable here: (I) the identity $H_{acc}=a K$ postulated for defining this transformation and (II) the correct flat limit in which we recover the genuine Rindler transformation since for $\omega \to 0$ we obtain the expansions
\begin{eqnarray}
T(t,x)&=&-\frac{1}{\omega}+x \sinh(a t)+{ O}(\omega)\,,\label{t1exp}\\
X(t,x)&=&x \cosh(a t) +{ O}(\omega)\,.
\end{eqnarray}     
Moreover, a nice surprise is the line element of the accelerated frame,
\begin{equation}
ds^2=a^2x^2\left(1-\frac{1}{4}\omega^2 x^2\right)dt^2+a\omega x^2 dt dx -dx^2\,,
\end{equation}
which in the flat limit becomes the well-known one (\ref{line}). Hereby we see that the transformations we propose here are no isometries - these change the metric. Note that the velocity with respect to the static frame of a mobile at rest in the origin  of the accelerated frame  is $V_0=\tanh (a t)$ as in the Rindler case. The difference is that this happens only in $x=0$ since in our case the velocities of the mobiles at rest in the accelerated frame depend on their positions, $x$, with respect to this frame. 

The chart $(t,x)$ corresponds to the following elegant parametrization,
\begin{eqnarray}
z^0&=&x \sinh(a t)+\frac{1}{2}\,\omega x^2\cosh(a t)\,,\\
z^1&=&x \cosh(a t)+\frac{1}{2}\,\omega x^2\sinh(a t)\,,\\
z^2&=&\frac{1}{\omega}-\frac{1}{2}\,\omega x^2\,,
\end{eqnarray}
of the de Sitter hyperboloid that can be obtained performing the boost transformation  in $M_3$, $(z^0,z^1,z^2)^T=b(t,x)(0,0,\frac{1}{\omega})^T$ where \cite{C2}
\begin{eqnarray}
&&b(t,x)=e^{-i a t k}e^{-i x p}\nonumber\\
&&=\left(
\begin{array}{ccc}
\cosh(at)& \sinh(a t)&0\\
\sinh(at)&\cosh(at)&0\\
0&0&1
\end{array}\right)\nonumber\\
&&\times \left(
\begin{array}{ccc}
1+\frac{1}{2}\omega^2x^2& \omega x&\frac{1}{2}\omega^2x^2\\
\omega x&1&\omega x\\
-\frac{1}{2}\omega^2x^2&-\omega x&1-\frac{1}{2}\omega^2x^2
\end{array}\right)\nonumber
\end{eqnarray}
This parametrization helps one to relate the coordinates of the chart $(t,x)$ to those of other charts we know as listed for example in Ref. \cite{Pascu}. Moreover, it shows how the accelerated frames can be defined in four-dimensional de Sitter spacetimes using Lorentz boosts along the acceleration direction. 

Our preliminary calculations indicate that our result seems to be  different from those obtained by other authors   \cite{BK1,BK2,BTW}. However, definitive conclusions could be drawn only after a complete study of the accelerated relative motion we defined here. 

\subsection*{Addendum}

This Addendum is written after we have found a more plausible new  solution of this problem \cite{Cprep} which for $a\to 0$ gives just the de Sitter metric. The crucial point is that therein  the quantum observables transform as
\begin{equation}\label{HHK}
H_{acc}=H+aK
\end{equation}
while here we considered that  $H_{acc}=aK$. We must observe that Eq, (\ref{HHK})  does make sense since it is plausible to add the energy of acceleration to the initial one instead of replacing it. Thus the conclusion could be that the metric proposed here remains a mere peculiar particular case (or a class room exercise)  with an obscure physical meaning.

\end{document}